# Bilayer Graphene on hexagonal Boron Nitride and the Family of quantum MetaMaterial

Takuya Iwasaki[1,*], Yoshifumi Morita[2,*]

[1] *Research Center for Materials Nanoarchitectonics, National Institute for Materials Science (NIMS), 1-1 Namiki, Tsukuba, Ibaraki 305-0044, Japan*
[2]*Faculty of Engineering, Gunma University, Kiryu, Gunma 376-8515, Japan*

*E-mail: IWASAKI.Takuya@nims.go.jp (T.I.), morita@gunma-u.ac.jp (Y.M.)

**Abstracts**

We report on the fabrication and characterization of dual-gated hexagonal boron nitride (hBN)/bilayer-graphene (BLG) superlattices. Due to the moiré effect, the hBN/BLG superlattice harbors an energy gap at the charge neutral point (CNP) and the satellites even without a perpendicular electric field. In BLG, moreover, the application of a perpendicular electric field tunes the energy gap, which contrasts with the single-layer graphene (SLG) and is linked to the family of rhombohedral multilayer graphene. Therefore, the hBN/BLG superlattice is accompanied with non-trivial energy-band topology and a narrow energy band with van Hove singularities. By the dual gating, systematic engineering of the energy-band structure can be performed and the carrier concentration is fine-tunable. This review is an extended version of the talk based on ref. [1], which is also a supplementary to the ref. T. Iwasaki, Y. Morita, K. Watanabe, T. Taniguchi, Phys. Rev. B106, 165134 (2022) and Phys. Rev. B109, 075409 (2024). The data show the universality and diversity in the physics of the hBN/BLG superlattices.

**Main**

**0. Introduction**

Carbon-based superlattices are promising from the viewpoint of low-dimensional quantum metamaterials. Here possible candidates for the templates of such superlattices are low-dimensional carbon materials like carbon nanotube [2] and graphene (Gr) [3]. When atomic-layer materials are stacked, a long-range spatial modulation can take place. It is called "moiré pattern" (Fig.1). In particular, the modulation by using hexagonal boron nitride (hBN) on graphene offers an unprecedented route to novel energy-band engineering (Fig.2,3) [4-7]. Applying such an engineering, we can study hBN/Gr superlattices by in-situ gate tuning. In BLG, in particular, interaction-driven symmetry breaking is expected due to its parabolic energy touching with a finite density of states, which is linked to the family of rhombohedral multilayer graphene. There the application of a perpendicular electric field tunes the energy gap in contrast to SLG [8]. Such systematic engineering of the energy-band structure and carrier-doping was demonstrated in ultra-clean dual-gated hBN/BLG superlattices with an alignment angle of nearly zero degrees [9].

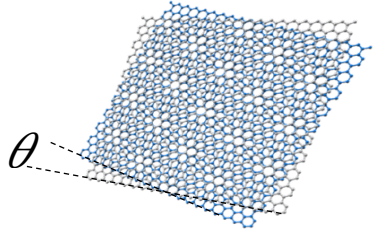

**Fig.1. "Moiré pattern".**

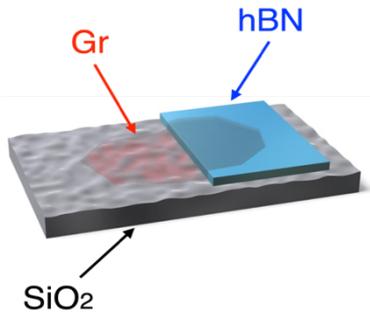 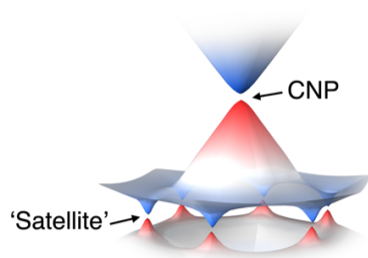

**Fig.2. Graphene (Gr) on hBN v.s. SiO$_2$.**   **Fig.3. Sketch for the energy band of hBN/Gr.**

Early days of the hBN/Gr started from the STM study. Please refer to the ref. [10] for a review on the early days of the hBN/Gr and the updated version [11]. Here in this paper, our focus is on hBN/BLG(/hBN/Graphite).

1. **hBN/Gr: "Lehrjahre"**

Graphene, a monolayer of carbon atoms, was discovered early in this century when it was exfoliated on SiO$_2$. This famous "Scotch-tape" method is now well known beyond the graphene community. Since every atom in graphene lives on the SiO$_2$ surface there, however, basic properties of graphene are highly-sensitive to the environment, i.e., the SiO$_2$ surface is not atomically flat with dangling bonds which are responsible for surface charge traps/charge-impurity doping (see Fig.2 for a cartoon image). Furthermore, the surface optical phonon has lower energy (compared with hBN), which causes difficulties in the device performance. A progress by the suspension of graphene improves the device quality (see ref. [12] for a benchmark). But the geometry of suspended graphene poses a severe limitation on the device architecture. At present, we are in the age of "after hBN" and graphene superlattice structures, hBN/Gr(/hBN), overcome such difficulties. The hBN is atomically flat and expected to be free of dangling bonds and surface charge traps (see Fig.2 for a cartoon image). Now, combined with the edge-contact technique discussed below, the stacking is basically decoupled from the metallization. As described in detail below, the graphene is only coupled to hBN in the fabrication process, which keeps the device away from scattering sources in the conventional process. But this is not the end of the story. Here we shall comment on two topics ("Edge Contact and Graphite BackGate") plus more on real settings for hBN/BLG(/hBN/Graphite).

### 1.1 Edge Contact and Graphite BackGate

In early days, electrical contact has been achieved by direct deposition of metals on the graphene surface ("surface contact"). However, graphene lacks surface bonding sites, which leads to large contact resistance. In search for better metal-graphene contact, we apply the graphene-metal "edge contact" [13], where the stacking is basically decoupled from the metallization. The "how-to" of this edge-contact technique is described below in real settings.

In spite of efforts described above, hopping transport between disorder-induced sub-gap states still remain, which causes some drawbacks in the device performance. By the use of back-gate under the hBN, the charged impurities in the $SiO_2$ are screened effectively to suppress the charge fluctuation. The graphite back-gate is now found to be more effective than other metallic ones which are associated with more interface contamination and possible grain boundaries. This point is further checked in a recent study, where tunable band-gap is confirmed in transport properties of hBN/BLG(/hBN/Graphite) [9]. It is well-known in optical experiments that the energy gap is tunable by the displacement field in BLG [8]. But it is only in recent days that ultra-high-quality devices are fabricated so that the electrical transport is sufficiently suppressed in the low-temperature regime that the data shows reasonable numeric of the estimated band gap. This was achieved by applying the graphite backgate.

### 1.2 More on real settings

See Fig. 4-8 for the fabrication details [14], where Fig. 8 is a summary for the main device which is characterized below. Exfoliated from bulk crystals by the "Scotch tape", BLG and hBN flakes were set on a $SiO_2$/Si substrate. Picking up such a flake by the edge of the polymer stamp and dropping it off very slowly in a controlled manner with an electrical motor, we assembled the hBN/BLG/hBN/Graphite heterostructure dubbed as "Stack", which was later (reactive ion-)etched into a Hall bar geometry combined with EB-lithography techniques. A graphite flake under the bottom hBN is used for the application of the back-gate voltage.

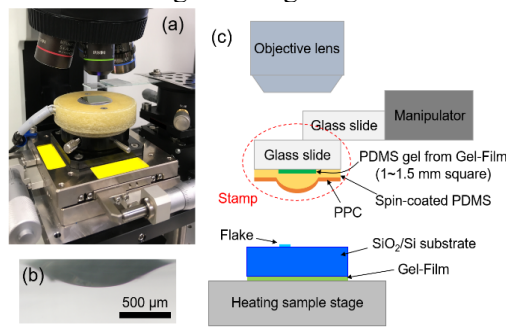

**Fig.4. Home-built dry-transfer equipment [14]. (a) Overall Picture. (b) Side-view of the Stamp. (c) Schematics.**

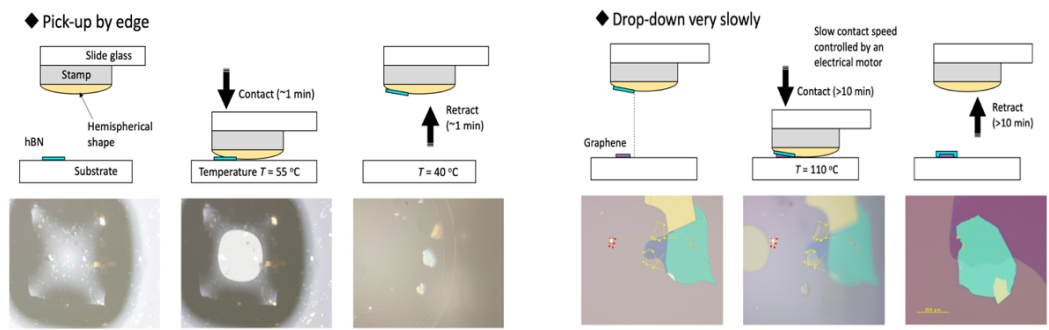

Fig.5. Pick-up and drop-down process with optical images. The scale bar shown in the last picture is 20.0μm (the first two is 25 times and the others are 5 times larger in scale compared with the last).

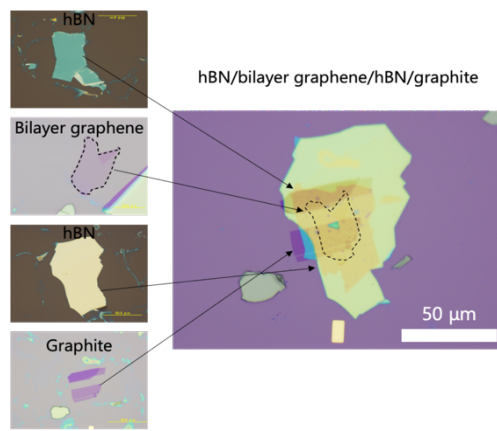

Fig.6. Optical image of the "Stack" (hBN/BLG/hBN/Graphite).

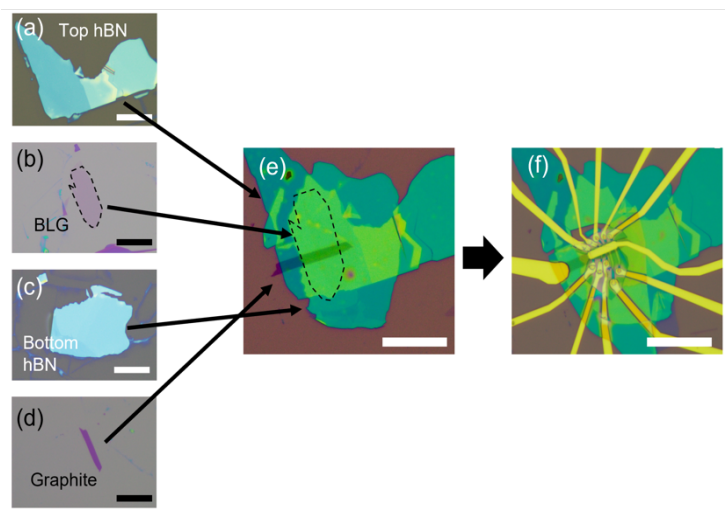

Fig.7. Optical images of the (a) top hBN, (b) BLG, (c) bottom hBN, (d) Graphite, (e) "Stack" (hBN/BLG/hBN/Graphite) and (f) fabricated device. The scale bar corresponds to 20 μm.

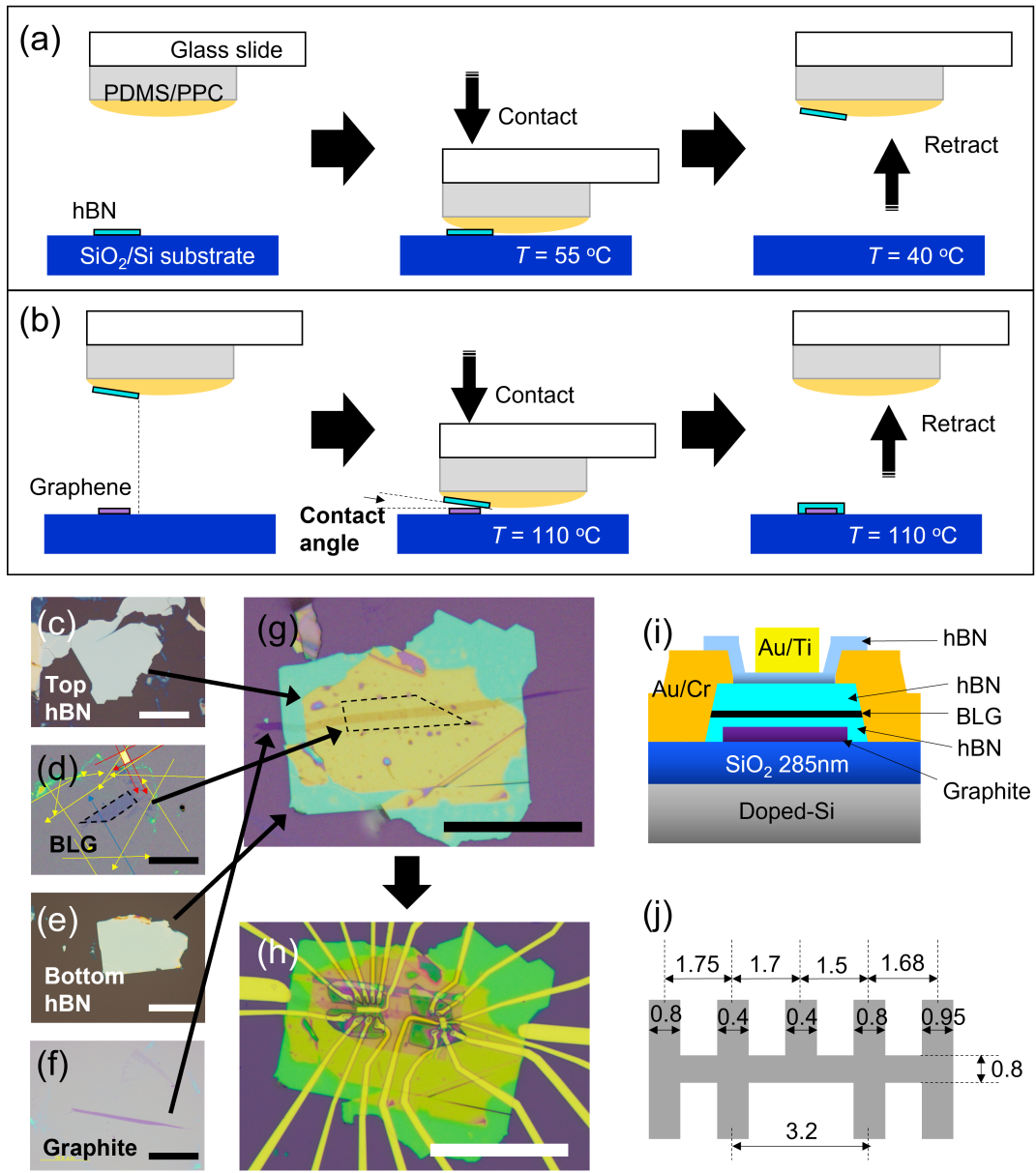

**Fig. 8.** Summary on the main device employed below. (a) Pick-up process for a hBN flake. (b) Drop-off process for a hBN flake onto a graphene flake. (c-h) Optical images of the (c) top hBN, (d) BLG (the edge is highlighted with the dashed line), (e) bottom hBN, (f) Graphite, (g) "Stack" and (h) fabricated device. The scale bar corresponds to 50 μm. (i) Cross-section of our device. (j) Channel geometry in our device. The scale is in μm units.

Finally in this subsection, more detail on the nano-fabrication process is given (Fig. 9, 10).

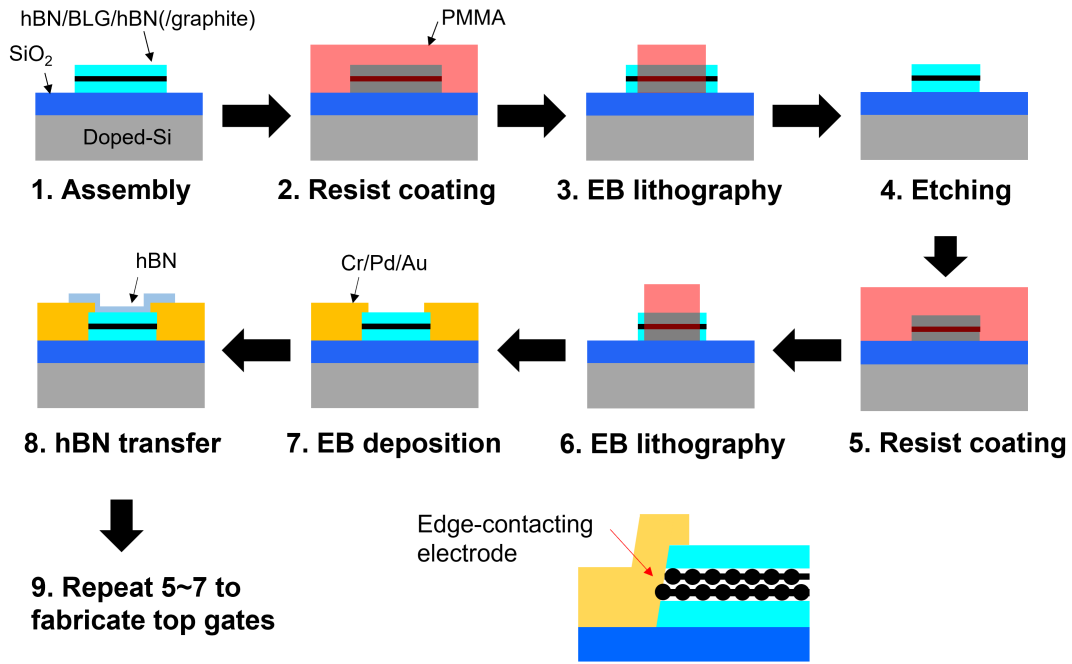

Fig.9 More detail on the nano-fabrication process.

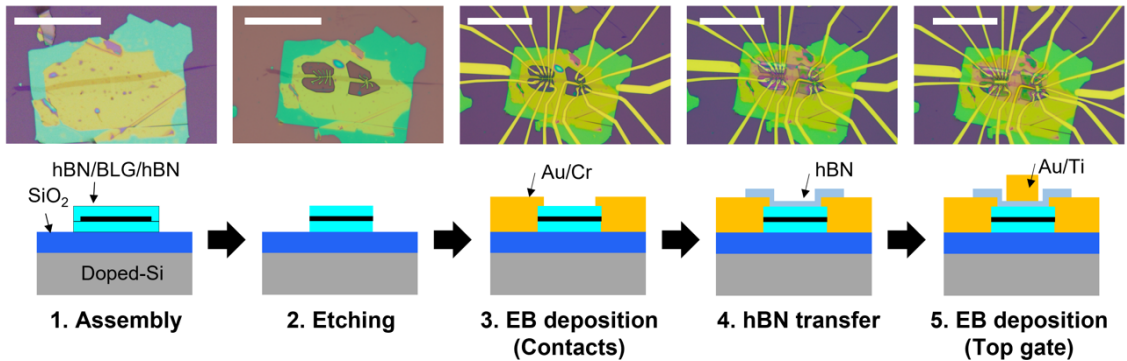

Fig.10 The upper and lower panels show the optical images and the schematic cross-section for each sequence of the nano-fabrication, respectively. The scale bars correspond to 50 μm.

Electrodes with one-dimensional edge contact [13] were also fabricated by an EB lithography, evaporation, deposition, and then another hBN layer was transferred onto the above device (Fig.9,10). The top gate (Au/ Ti) was fabricated in the same manner as the electrical contact and then we got a final form.　A comment is in order for experimental outputs in the following. A four-terminal method with AC lock-in techniques (an excitation current $I \sim$ 10-100 nA and a frequency of 17 Hz) was applied for the longitudinal resistivity $\rho_{xx} = (V_{xx}/I)(W/L)$ and the Hall resistivity $\rho_{xy} = V_{xy}/I$, where $L$ and $W$ are the channel length and width, respectively, $V_{xx}$ and $V_{xy}$ are the voltage measured between the Hall-bar electrodes. The device was set at low temperature ($T$) in a $^4$He cryostat with a superconducting magnet with a perpendicular magnetic field ($B$). The optical image for the main device employed below is shown in Fig.11.

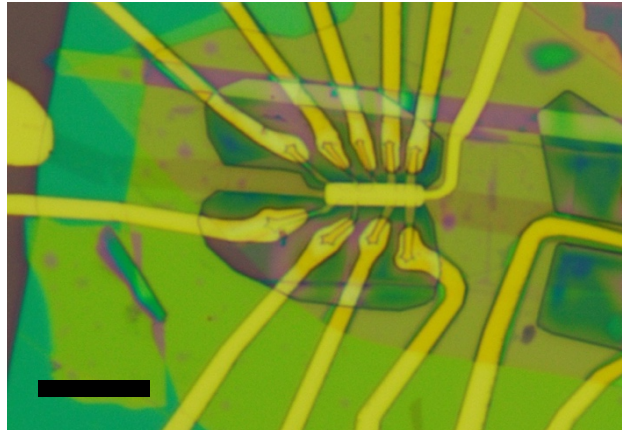

**Fig.11 Optical image of the main device (see also Fig. 13). The scale bar corresponds to 10 μm.**

2. **hBN/BLG: from "the Beginning" to "Dual gating"**

An "evolution" of our devices is shown in Fig.12 (2019 ver.) and Fig.13 (2022 ver.).

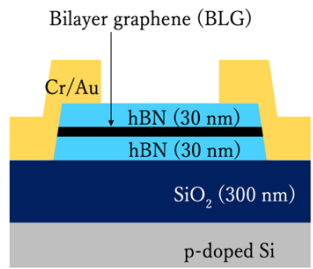

**Fig.12 Schematics of an ultra-clean hBN/BLG device (2019 ver.) with an alignment angle of nearly zero degrees, whose cross-section is shown above. This was employed in the ref. [18] etc. (please check also, for example, arXiv: 1901.09356).**

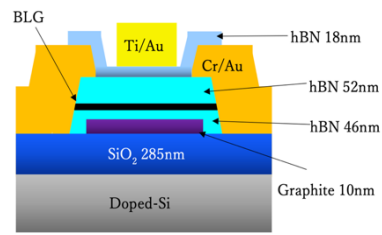

**Fig.13 Schematics of an ultra-clean "dual-gated" hBN/BLG device (2022 ver.) with an alignment angle of nearly zero degrees, whose cross-section is shown above. See also Fig.11. This is the main device employed in the following.**

## 3. Global phase diagram of hBN/BLG

In the following, the experiment is carried out in the "Stack" (hBN/BLG/hBN/Graphite, which is also called hBN/BLG for simplicity) with top and bottom gates (so called "dual- gated structure"), where hBN plays the role of high-quality dielectric on both sides and we can independently control the top gate voltage ($V_{tg}$) and the bottom gate ($V_{bg}$), i.e., the carrier density $n$ and the perpendicular displacement field $D$ (Fig.14,15).

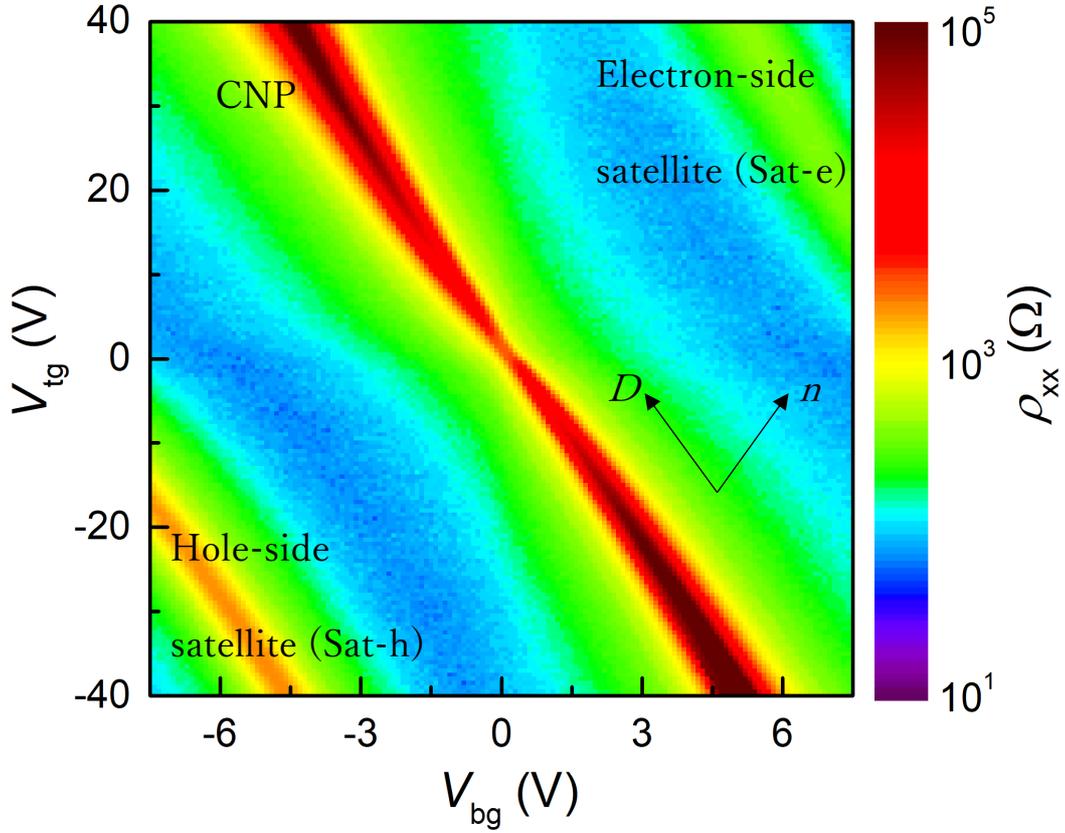

**Fig. 14. Intensity map of the longitudinal resistivity $\rho_{xx}$ as a function of $V_{tg}$ and $V_{bg}$, i.e., the displacement field $D$ and the carrier density $n$ at $T$ = 1.6 K, $B$ = 0 T [9].**

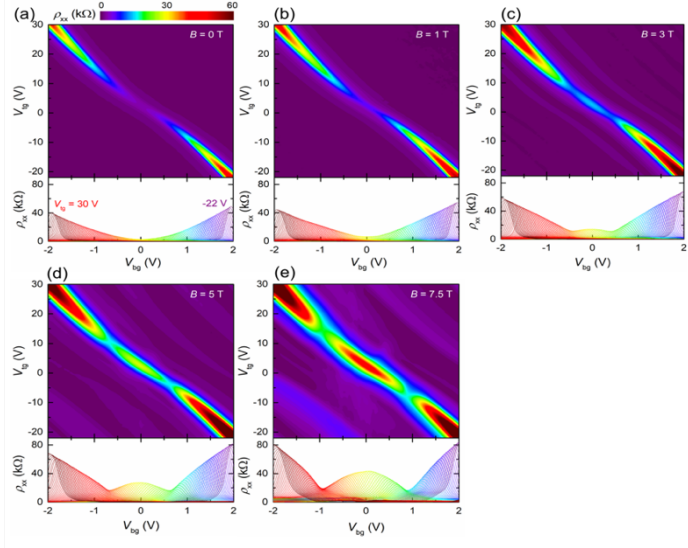

**Fig.15. Magnetotransport near the CNP [9].** The upper panels show the mapping of $\rho_{xx}$ on the $V_{tg}$-$V_{bg}$ plane at $T$ = 1.6 K. The lower panels show $\rho_{xx}$ v.s. $V_{bg}$ for various $V_{tg}$. (a) $B$ = 0 T, (b) 1 T, (c) 3 T, (d) 5 T, and (e) 7.5 T.

### 3.1 Role of flavor at the CNP
#### 3.1.1 An intro: quantum flavor currents

A valley is a "hidden" degree of freedom built into electrons for several solid-state systems [15]. Here the valley(-spin) degree of freedom is also called "flavor". Gapped Dirac materials with a broken inversion symmetry can exhibit topological flavor current even without a broken time-reversal symmetry. Such a topological flavor current was first demonstrated for hBN/SLG superlattices [16, 17] and also observed in hBN/BLG superlattices [18]. In the stacking of hBN on/below BLG, a slight mismatch in lattice constants can cause the "moiré effect" and the hBN/BLG superlattice harbors an energy gap (Fig. 3). Therefore, hBN/BLG is accompanied with non-trivial energy-band topology, i.e., accumulated Berry curvature and a narrow energy band with van Hove singularities. Here a key issue in this topic is the concept of *quantum limit*, which is considered to hold for $\sigma_{xx} \ll \sigma_{xy}^{v}$ (the valley Hall angle above ~1) in ultra-clean devices, and the emergence of the quantum valley Hall state (qVHS), which is associated with *non-local* resistance of the order $\sim h/e^2$ [17]. This is a topic which needs another review and we do not go further here. See ref. [19] for a recent status in the context of dual-gated hBN/BLG superlattices. Here we only note that this state should play the role of "parent state", which bears competing (topological) orders away from the CNP.

### 3.1.2 Flavor symmetry breaking at the CNP

Fig. 16 shows an intensity map of the longitudinal resistivity $\rho_{xx}$ as a function of the displacement field, $D$ and the magnetic field, $B$ (both $D$ and $B$ are applied perpendicular to the substrate), at 1.6 K. In these superlattices, a long wavelength moiré pattern leads to an energy gap at the charge neutral point (CNP) at $n = 0$ cm$^{-2}$. As shown in Fig. 16, under high magnetic fields, the resistance peak shows two "conductance spikes/resistance dips", which imply quantum phase transitions with "flavor" (spin/valley)-symmetry breaking as discussed in ref. [9].

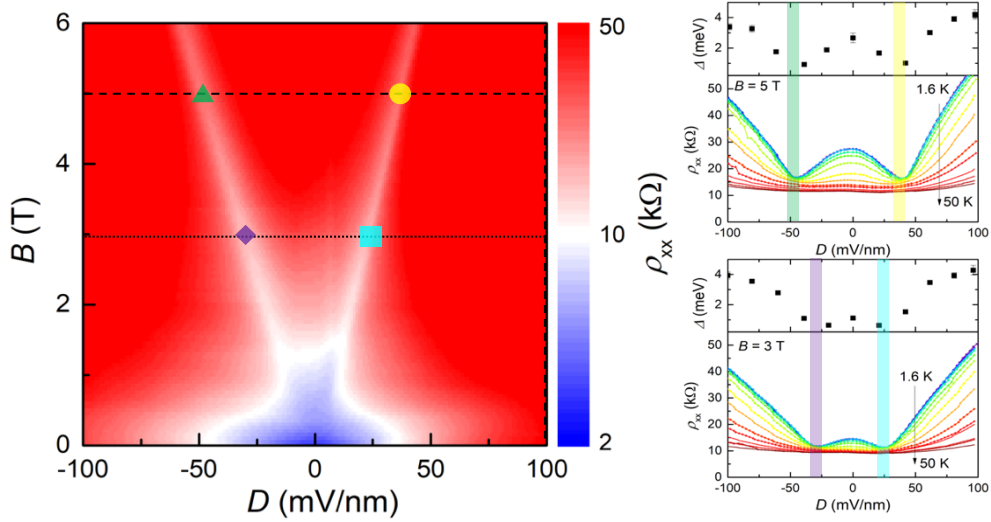

**Fig. 16. (Left Panel) Phase Diagram at the CNP [9]. The $\rho_{xx}$ is shown as a function of a displacement field $D$ and a magnetic field $B$ at 1.6K. (Right Panel) The gap $\Delta$ and the $\rho_{xx}$ as a function of $D$ with varying temperature, where the colored bar corresponds to the point with the same color in the left panel.**

### 3.2 Beyond the CNP: competing orders in the "Darwinian Sea"
#### 3.2.1 An intro: "Butterfly"

Here let us start form the magneto-resistance (MR) study away from the CNP. The quantum Hall effects (QHE) in hBN/Gr is also called "Butterfly" in this review (Fig.17,18). This topic, especially in the "integer" regime, is now well-established and, for reviews on the QHE in hBN/Gr, please refer to ref.'s [10,11]. More generally, see for the MR study in hBN/Gr, e.g., ref. [9, 20] and references therein. The MR can be a sensitive probe of competing orders.

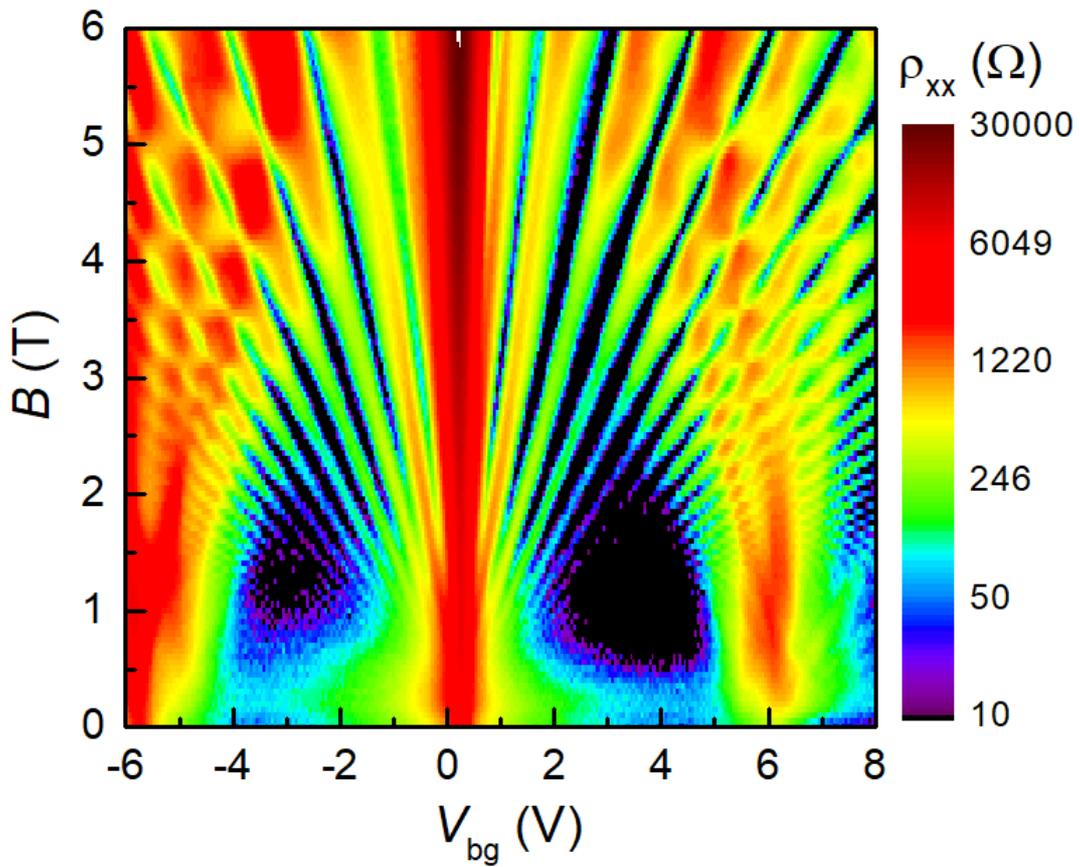

Fig. 17. QHE in our device (1.6K), which shows the "Butterfly".

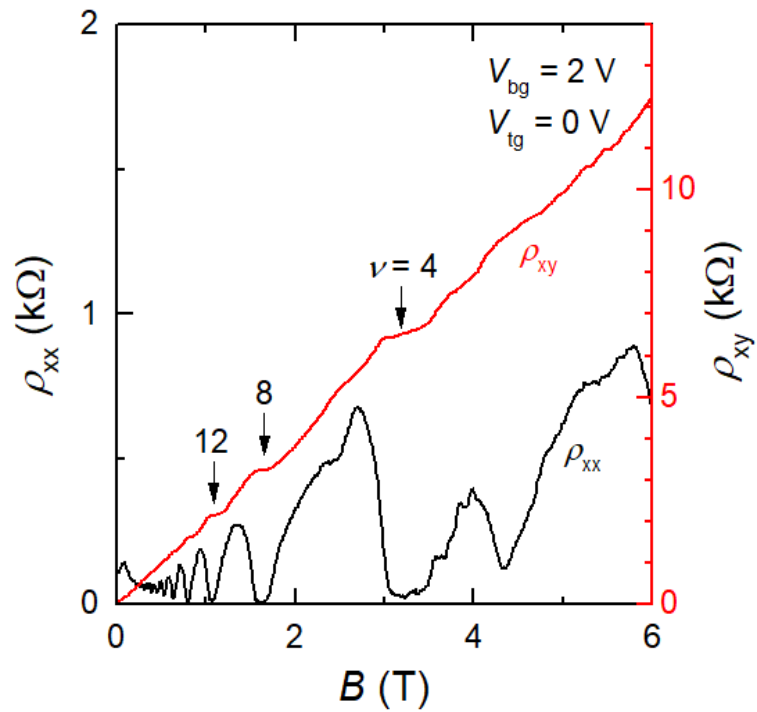

Fig. 18. QHE in another device (1.7K), which shows typical "BLG" plateaus.

### 3.2.2 Competing Orders: "Darwinian Sea"

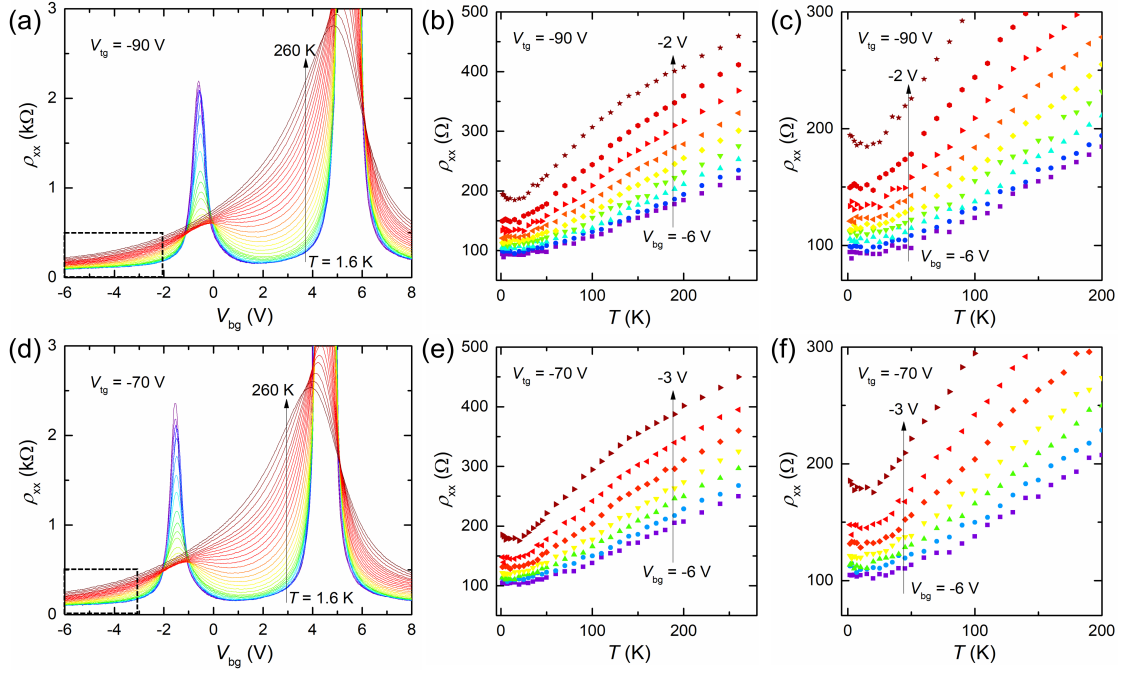

**Fig. 19.** Transport properties away from the CNP [9]. (a) $\rho_{xx}$ as a function of $V_{bg}$ with $V_{tg}$ = -90 V at $T$ = 1.6 K, $B$ = 0 T. (b) $\rho_{xx}$ as a function of $T$ for various $V_{bg}$ (corresponding to the dashed box region in (a)) with $V_{tg}$ = -90 V. (c) Zoom up of (b). (d-f) Same plots as (a-c) with $V_{tg}$ = -70 V.

See Fig. 19. The two resistance peaks of the CNP and the satellite are shown in Fig.19 (a,d). Next move away from there. As is well established, the resistance-temperature (*R-T*) characteristics are basically due to phonon contribution in the graphene family [21]. On the other hand, some departure from such an assignment is also shown in Fig. 19, which shows an insulating behavior (combined with a "resistance bump") in the low-temperature limit and contrasts with conventional phonon contribution. Although this type of insulating behavior has also been observed in other graphene devices, the origin has not been identified yet in a definite manner and this has been routinely called "correlated insulator/semimetal" (or "Mott") in several literatures. "Hidden order" like (unconventional) charge/spin/valley density wave is a possible candidate for the origin of such departures [22], but more careful assignment is left as future works. Another physics like the "fractional" one can also be competing with other orders. More generally, such phenomena with a resistance bump (and/or drop nearby) can be due to various broken symmetries or topological order by many-body effects [23, 24]. Such orders are always competing in the narrow energy band with van Hove singularities ("Darwinian Sea") and there are several ways to fix the survivor and "resolve the singularities". See also Fig. 20 for more on the resistance bump and/or drop in a different device.

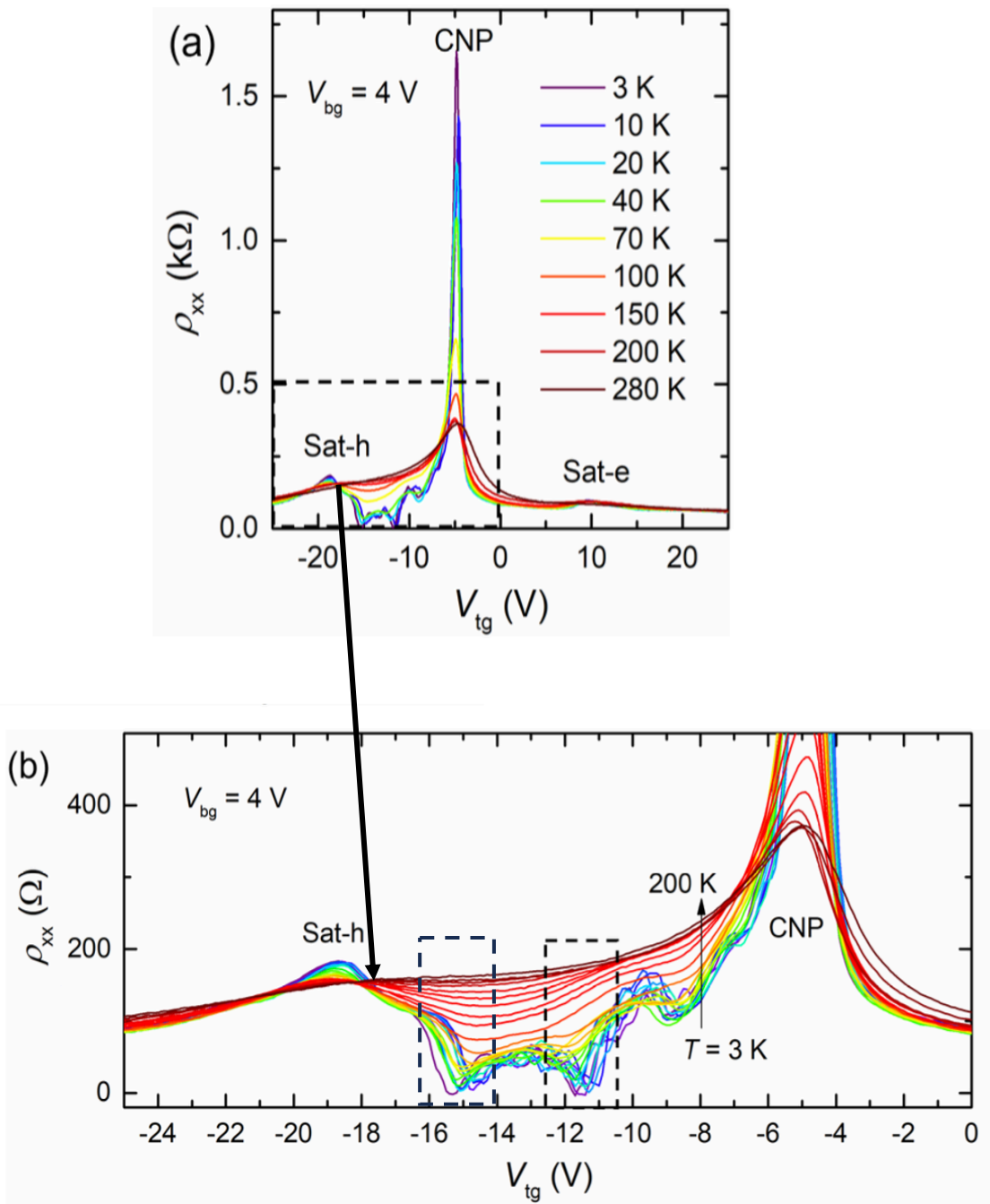

Fig. 20. Characterization of another device at $B = 0$ T [9]. (a) $\rho_{xx}$ as a function of $V_{tg}$ for various $T$ at $V_{bg} = 4$ V. (b) Zoom up of the dashed box in (a). CNP and Sat-h/e correspond to the charge neutral point and the satellites on the hole and electron side respectively.

4. Novel quantum MetaMaterial: hBN/BLG & The Family

As we have discussed above, BLG on hBN has so much potential. The concept of "superlattice" can be traced back to the celebrated works by L. Esaki and R. Tsu [25] and HEMT [26], which is a stage for QHE. Based on the superlattice "hBN/BLG", advances in nanofabrication techniques should make it possible to make state-of-the-art devices with single electron transistor (SET) as shown in Fig. 21, quantum point contacts (QPC) and more including their "aufheben". Here note that the QPC is a basic building block for the "fermionic quantum optics", which contrasts with the "single electronics" in SET. Although such projects (including our collaborators' [27]) are done/on-going in SLG and BLG (without alignment to hBN with nearly zero degress) [28-30], not so much is done in the context of hBN/BLG [31] and more is left to be studied in the future to come, e.g., electrical switching between the competing orders including a topological one. There, after the Wanderjahre described in this review, the low-dimensional/nano- physics should open a door to the next level combined with the carbon physics [1, 2, 32].

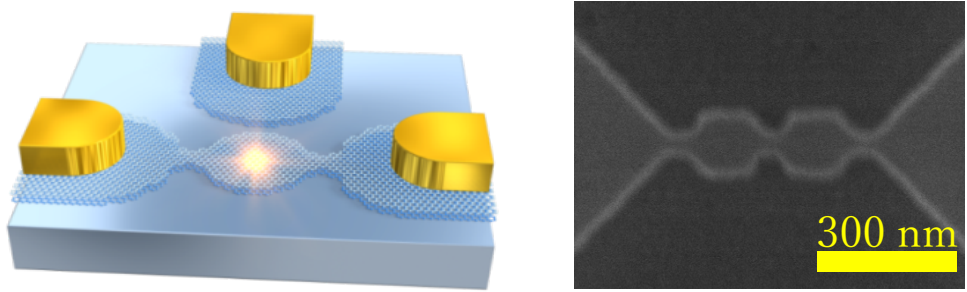

**Fig. 21. Graphene SET: Schematics v.s. SEM image of a hBN/BLG SET [31].**

In summary, we have discussed dual-gated hBN/BLG superlattices, where both the energy-band structure and the carrier concentration are tunable by the dual gating. Several transport properties are revealed and this review should lay the basis for the "global phase diagram" of the dual-gated hBN/BLG superlattices and beyond, e.g., fabrication and study of nanostructures like SET, QPC and Josephson Junctions and their integrated structures in the "quantum circuit".

**Acknowledgements**
We thank all our collaborators, although we do not list them up all. In particular, T. Taniguchi and K. Watanabe played a crucial role for providing us ultra-pure hBN crystals. We also thank H. Osato, E. Watanabe, and D. Tsuya from the NIMS Nanofabrication Facility. This work was partially supported by JPSJ KAKENHI Grant No. 21H01400, and NIMS Nanofabrication Facility.

**References**

[1] Y. Morita, T. Iwasaki, K. Watanabe, T. Taniguchi, Extended Abstracts of the 2023 International Conference on Solid State Devices and Materials (SSDM2023), Nagoya, pp61 (2023).

[2] S. Iijima, Nature 354, 56 (1991).

[3] K. S. Novoselov et al., Science 306, 666 (2004); K. S. Novoselov et al., Nature 438, 197 (2005); A. K. Geim, K. S. Novoselov, Nat. Mater. 6, 183 (2007).

[4] R. Bistritzer, A. H. MacDonald, Proc. Natl. Acad. Sci. U.S.A. 108, 12233 (2011); Phys. Rev. B 84, 035440 (2011).

[5] M. Yankowitz et al., Nature Physics 8, 382 (2012).

[6] K. Watanabe, T. Taniguchi, H. Kanda, Nat. Mater. 3, 404 (2004).

[7] C. R. Dean et al., Nat. Nanotechnol. 5, 722 (2010).

[8] Y. Zhang et al., Nature 459, 820 (2009).

[9] T. Iwasaki, Y. Morita, K. Watanabe, T. Taniguchi, Phys. Rev. B106, 165134 (2022).

[10] M. Yankowitz et al., J Phys Condens Matter, 26, 303201 (2014).

[11] M. Yankowitz et al., Nat. Rev. Phys., 1, 112 (2019).

[12] K. I. Bolotin et al., Solid State Communications, 146, 351 (2008).

[13] L. Wang et al., Science 342, 614 (2013).

[14] T. Iwasaki et al., ACS Appl. Mater. Interfaces 12, 7 (2020).

[15] D. Xiao, W. Yao, Q. Niu, Phys. Rev. Lett. 99, 236809 (2007).

[16] R. V. Gorbachev et al., Science 346, 448 (2014).

[17] K. Komatsu et al., Sci. Adv. 4, eaaq0194 (2018).

[18] K. Endo et al., Appl. Phys. Lett. 114, 243105 (2019).

[19] T. Iwasaki, Y. Morita, K. Watanabe, T. Taniguchi, Phys. Rev. B109, 075409 (2024).

[20] T. Iwasaki et al., Carbon 175, 87 (2021); Scientific Reports 11, 18845 (2021).

[21] E. H. Hwang, S. D. Sarma, Phys. Rev. B77, 115449 (2008); S. Das Sarma et al., Rev. Mod. Phys. 83, 407 (2011).

[22] T. Iwasaki et al., Appl. Phys. Express 13, 035003 (2020); arXiv:2110.00510 (, which is based on T. Iwasaki's talk at SSDM2021/Ext. Abstr. Solid State Devices and Materials, 2021, p. 445, and plays the role of a supplementary to the APEX paper).

[23] R. Nandkishore, L. Levitov, arXiv:1002.1966; Phys. Rev. B82, 115124 (2010); Phys. Rev. Lett. 104, 156803 (2010); Z. Dong, M. Davydova, O. Ogunnaike, L. Levitov, Phys. Rev. B 107, 075108 (2023) ; Z. Dong, M. Davydova, O. Ogunnaike, L. Levitov, Phys. Rev. B 107, 174512 (2023).

[24] F. Wu, A. H. Macdonald, I. Martin, Phys. Rev. Lett. 121, 257001 (2018).

[25] L. Esaki, R. Tsu, IBM Journal of Research and Development 14, 61 (1970).

[26] T. Mimura et al., Jpn. J. Appl. Phys. 19, 225 (1980).

[27] N. F. Ahmad et al., Appl. Phys. Lett. 114, 023101 (2019); Scientific Reports 9, 3031 (2019).

[28] L. Veyrat et al., Nano Lett. 19, 635 (2019).

[29] M. Eich et al., Phys. Rev. X 8, 031023 (2018).

[30] L. Banszerus et al., Nano Lett. 18, 4785-4790 (2018).

[31] T. Iwasaki et al., Nano Lett. 20, 2551-2557 (2020).

[32] M. S. Dresselhaus, G. Dresselhaus, Advances in Physics 30, 139 (1981); R. P. Smith et al., Physica C514, 50 (2015).